\documentclass[showpacs,preprintnumbers,amsmath,amssymb,prl,twocolumn]{revtex4}

\usepackage{graphicx}
\usepackage{dcolumn}
\usepackage{bm}
\def\comment#1{}

\begin{document}

\title{Quantum electrodynamics in $2+1$ dimensions, confinement, and the 
stability of $U(1)$ spin liquids}
\author{Flavio S. Nogueira}
\affiliation{Institut f{\"u}r Theoretische Physik,
Freie Universit{\"a}t Berlin, Arnimallee 14, D-14195 Berlin, Germany}
\author{Hagen Kleinert}
\affiliation{Institut f{\"u}r Theoretische Physik,
Freie Universit{\"a}t Berlin, Arnimallee 14, D-14195 Berlin, Germany}

\date{Received \today}

\begin{abstract}
Compact quantum electrodynamics in $2+1$ dimensions often arises as  
an effective theory for a Mott insulator, with the Dirac fermions 
representing the low-energy 
spinons. An important and controversial issue in this context 
is whether a deconfinement transition takes place. We perform a  
renormalization group analysis to show that deconfinement occurs 
when $N>N_c=36/\pi^3\approx 1.161$, where $N$ is the number 
of fermion replica. For $N<N_c$, however, there are 
two stable fixed points separated by a 
line containing a unstable non-trivial fixed point: a fixed point 
corresponding to the scaling limit of the non-compact theory, and 
another one governing the scaling behavior of the compact theory.    
The string tension associated to 
the confining interspinon potential is shown to exhibit a universal jump as $N\to N_c^-$. 
Our results imply the stability of a spin liquid at the 
physical value $N=2$ for Mott insulators. 
\end{abstract}

\maketitle

An important topic currently under discussion in condensed matter 
physics community is the emergence of deconfined quantum critical 
points in gauge theories of Mott insulators in $2+1$ dimensions 
\cite{Sachdev,Senthil}. A closely related problem concerns the 
stability of $U(1)$ spin liquids in $2+1$ dimensions 
\cite{Herbut1,Hermele}. In either case,  
models which are often considered as 
toy models in the high-energy physics literature are supposed 
to describe the low-energy properties of real systems 
in condensed matter physics. For instance, 
a model that frequently appears in the condensed matter 
literature is the $(2+1)$-dimensional quantum electrodynamics 
(QED3) \cite{Pisarski,Appelquist}. 
It emerges, for instance, as an effective theory for Mott insulators 
\cite{Affleck,Marston,Kim}. Let us briefly recall how it arises in this context. 
The Hamiltonian of a $SU(N)$ Heisenberg antiferromagnet is written in a slave-fermion 
representation as 
$H=-(J/N)\sum_{\langle i,j\rangle}f_{i\alpha}^\dagger f_{j\alpha}f_{j\beta}^\dagger f_{i\beta}$, 
where the local constraint $f_{i\alpha}^\dagger f_{i\alpha}=N/2$ holds. A Hubbard-Stratonovich 
transformation introduces the auxiliary field $\chi_{ij}=\langle f_{i\alpha}^\dagger f_{j\alpha}\rangle$ 
\cite{Affleck}. The resulting effective theory can be treated 
as a {\it lattice gauge theory}, where the gauge field $A_{ij}$ emerges as the phase of 
$\chi_{ij}$, i.e., $\chi_{ij}=\chi_0\mathrm{e}^{\mathrm{i}A_{ij}}$, 
where $\chi_0$ is determined from mean-field theory. The 
$(2+1)$-dimensional low-energy effective 
Lagrangian in imaginary time has the form \cite{Hermele,Affleck,Marston,Kim}
\begin{equation}
{\cal L}=\frac{1}{4e_0^2}F_{\mu\nu}^2+\sum_{a=1}^N\bar \psi_a\gamma_\mu(\partial_\mu+iA_\mu)\psi_a,
\end{equation}
where each $\psi_a$ is a four-component Dirac spinor and $F_{\mu\nu}=\partial_\mu A_\nu-\partial_\nu A_\mu$ 
is the usual field strength tensor. A rough estimate of the bare gauge coupling is given by 
$e_0^2\sim\chi_0^4a^3$, where $a$ is the lattice spacing.  

An anisotropic version of QED3 has  
also been studied in the context of phase fluctuations in $d$-wave 
superconductors \cite{Tesanovic,Herbut2}. A key feature of the  
QED3 theory of Mott insulators is its parity conservation. 
In fact, it is possible to introduce two different QED3s, one 
which conserves parity and one which does not. The latter theory involves 
two-component spinors, and allows for a {\it chirally-invariant} mass term  
which is not parity-invariant. In such a QED3 theory a   
Chern-Simons term \cite{Deser} is generated by fluctuations  
\cite{Niemi}. The QED3 theory relevant to Mott insulators and 
$d$-wave superconductors involves four-component spinors, 
and does possess chiral symmetry 
\cite{Pisarski,Appelquist}. In such a model, the chiral 
symmetry can be spontaneously broken through the dynamical 
generation of a fermion mass. In the context of Mott insulators,  
the chiral symmetry breaking corresponds to the 
development of N{\'e}el order \cite{Kim}. 
Indeed, the nonzero condensate $\langle\bar{\psi}\psi\rangle$ corresponds to the 
staggered magnetization. 
Since in condensed matter 
physics parity-conserving QED3 is not just a toy model, 
and that the 
four-component Dirac spinors represent physical excitations --- the low-energy spinons, we  
may call this theory  
{\it quantum spinodynamics}.

An important feature 
of QED3 for Mott insulators is that the $U(1)$ gauge 
group is {\it compact}. The compactness causes important changes in 
the physical properties of the theory. 
It allows for quantum excitation of 
magnetic monopoles which play an important role in determining 
the phase structure of the theory. This has been known  
for a long time. In particular, Polyakov \cite{Polyakov} 
has shown that compact Maxwell theory in $2+1$ 
dimensions confines permanently electric test charges.  
The electrostatic potential has the form 
$V(R)\sim R$, instead of the usual two-dimensional Coulomb potential 
$V(R)\sim\ln R$ of the non-compact Maxwell theory in $2+1$ 
dimensions. Since $V(R)\sim R$ holds for all values of the 
gauge coupling, the compact $(2+1)$-dimensional Maxwell theory does 
not exhibit any phase transition, i.e., the confinement is permanent. 
This theory is equivalent to a Coulomb gas of magnetic monopoles 
in three dimensions and it is well known that such a 
gas does not undergo any phase transition. However, 
when matter fields are included the situation changes, and 
a deconfinement transition may occur. Indeed, matter fields induce shape fluctuations 
of the {\it electric} flux tube, leading to a 
correction term to the linearly confining potential. Using a string 
model for the electric 
flux tube, L\"uscher \cite{Symanzik} found  
\begin{equation}
\label{string}
V(R)=\sigma R - \frac{(d-2)\pi}{24R}+{\cal O}(1/R^2),
\end{equation}
where $\sigma$ is the string tension.
At the 
deconfining critical point, the string tension vanishes and only 
the L\"uscher term remains at long distances. 
It 
represents the ``black-body'' energy of the 
$(d-2)$ transverse 
fluctuations of the two-dimensional worldsheet of the string \cite{Note3}. 

Interestingly, by studying  QCD near four dimensions, 
Peskin \cite{Peskin} 
found that, at the critical point, the 
interquark potential does have the $1/R$-behavior for all $d\in(4,4+\epsilon)$, 
and argued that this should also be valid outside this small dimension interval.  
Recently we \cite{KNS} have found that such a behavior is also realized 
in an Abelian U(1)-gauge theory  for $d\in(2,4)$, provided that this is coupled to matter fields. 

In order to better illustrate this  
mechanism, we shall explicitly perform the calculation of the 
interspinon potential to one-loop order in arbitrary space-time dimension $2<d\leq 4$, 
going to the   
case of interest $d=3$ at the end \cite{Note1}. The 
potential is defined by
\begin{equation}
\label{explicitV}
V(R)=-e_0^2\int\frac{d^{d-1} q}{(2\pi)^{d-1}}
\frac{e^{i{\bf q}\cdot{\bf R}}}{{\bf q}^2[1+\Pi({\bf q})]},
\end{equation}
where $e_0$ is the bare electric charge, and 
$\Pi(q)$ is the vacuum polarization. At one-loop order,  
the vaccum polarization is given by $\Pi(q)=8A(d)Ne_0^2|q|^{d-4}$,
with  $A(d)=\Gamma(2-d/2)\Gamma^2(d/2)/[(4\pi)^{d/2}\Gamma(d)]$. 
At large distances the vacuum polarization gives the more relevant 
contribution to the scaling behavior if  
$2<d<4$, and the interspinon potential is given by 
\begin{equation}
\label{Vcrit}
V(R)=-\frac{1}{2^{d+1}\pi^{(d-2)/2}\Gamma(d/2-1)A(d)N}\frac{1}{R}.
\end{equation}
For $d=3$ the above potential becomes simply 
$V(R)=-4/(\pi NR)$. Interestingly, by 
expanding (\ref{Vcrit}) near $d=2$, we obtain at lowest order
\begin{equation}
\label{Vcrit1}
V(R)\approx-\frac{(d-2)\pi}{8NR},
\end{equation}
which has for $N=3$ replica precisely the form of the L\"uscher term, although 
this theory has no confinement. In order to allow for this, we
compatify the U(1) gauge group
which gives rise to 
magnetic monopoles. In the absence of fermions this 
theory in $d=3$ is known to be described via a duality transformation  
by the sine-Gordon Lagrangian \cite{Polyakov}
\begin{equation}
\label{SG}
{\cal L}=\frac{1}{2}\left(\frac{e_0}{2\pi}\right)^2
(\partial_\mu\chi)^2-2z_0\cos\chi.
\end{equation}
This is also the field theory of a Coulomb gas of monopoles, 
with $z_0$ being the bare fugacity of the gas. The RG equations for 
a Coulomb gas in $d=3$ were obtained a long time ago by Kosterlitz 
\cite{Kosterlitz}. His results can be used here to obtain the RG 
equation for the gauge coupling in the absence of fermionic matter. 
By introducing the dimensionless couplings 
$f\equiv e^2(l)/e_0^2$ and $y\equiv z(l)/(e_0^2)^3$, where 
$l=\ln(e_0^2r)$ is a logarithmic length scale, we 
obtain
\begin{eqnarray}
\label{betaf-Polyakov}
\frac{df}{dl}&=&4\pi^2y^2+f,\\
\label{betay}
\frac{dy}{dl}&=&\left(3-\frac{\pi^3}{f}\right)y.
\end{eqnarray} 
The above equations imply that there is no fixed point for the gauge coupling. 
Therefore, the compact three-dimensional Maxwell theory does not undergo 
any phase transition. The photon mass $M^2=8\pi^2z/e^2$ is always 
non-zero and the theory confines permanently the electric charges. 
It can be seen from Eq. (\ref{betaf-Polyakov}) that a kind of 
{\it anti-screening} happens in this theory, which is responsible 
for confinement. Indeed, we can rewrite 
Eq. (\ref{betaf-Polyakov}) as $df/dl=(1-\hat{\gamma}_A)f$, with 
$\hat{\gamma}_A=-4\pi^2y^2/f$. The negative sign of 
$\hat{\gamma}_A$ is actually a remarkable example of 
the intimate link between asymptotic 
freedom and confinement \cite{Note4}. 

Next we obtain the 
modification to Eqs. (\ref{betaf-Polyakov}) and (\ref{betay})  
due to the coupling with  
the matter fields. 
In order to derive the RG equations including matter we have 
employed a formalism similar to the one developed by Young 
\cite{Young} in the case of the two-dimensional Coulomb gas. 
This formalism is based on a 
mean-field self-consistent approximation and 
applies very well to the $d>2$ case, since $d=2$ 
is the upper critical dimension for the Coulomb gas. The needed 
modification comes from the extra renormalization of the gauge coupling 
due to the vacuum polarization. This leads to an effective 
charge $e^2(l)=\varepsilon(e^l/e_0^2)Z_A(l)e^l e_0^2$, where 
$\varepsilon(r)$ is the scale-dependent ``dielectric'' constant 
of the Coulomb gas of magnetic monopoles, and 
$Z_A(l)$ is the gauge field wave function 
renormalization. In terms of dimensionless 
couplings this leads to        
\begin{equation}
\label{betaf-new}
\frac{df}{dl}=4\pi^2y^2+(1-\gamma_A)f,
\end{equation}
where $\gamma_A\equiv -d\ln Z_A/dl$. 
However, the expression of $y^2$ 
in terms of bare variables is different from before, being given 
by $y^2=(32\pi^2/3)z_0^2Z_A(l) e^{6l-u(l)}/(e_0^2)^6$, where $u(l)\equiv U(e^l/e_0^2)$ is 
a self-consistent magnetic monopole potential  
satisfying $du/dl=\pi^3/f$ \cite{Calcs}.
Therefore, the coupling to matter modify also Eq. (\ref{betay}) to
\begin{equation}
\label{betay-new}
\frac{dy}{dl}=\left(3-\frac{\pi^3}{f}-\frac{\gamma_A}{2}\right)y.
\end{equation}
Note the crucial difference between the analysis made here and the 
one of Refs. \cite{KNS}, \cite{KNS1} and \cite{Herbut3}. 
{\it There it was assumed that the 
underlying non-compact  
theory is critical, and  monopoles were 
introduced only at that point. This corresponds to take the RG function $\gamma_A$
at the fixed point of the non-compact theory, i.e., $\gamma_A^*\equiv\eta_A=1$} 
\cite{etaArefs} {\it for $d=3$}. 
In this way, Eqs. (\ref{betaf-new}) and (\ref{betay-new}) become 
similar to the RG equations of a Kosterlitz-Thouless phase transition 
\cite{KT}, except that the present dimensionality is three instead of two 
\cite{KNS1,KSN}. Our Eqs. 
(\ref{betaf-new}) and (\ref{betay-new}) have 
the advantage of being valid at all length scales. Eqs. (\ref{betaf-new}) and 
(\ref{betay-new}) are similar to the ones in the work of Hermele {\it et al.} 
\cite{Hermele}. There is, however, an important difference: Eq. (\ref{betay-new}) 
contains the correction proportional to $\gamma_A$ which is absent in Ref. \cite{Hermele}. 
This will allow us to strenghten considerably the results obtained by these authors. 
It is important to emphasize that the additional term in Eq. (\ref{betay-new}) {\it cannot} 
be neglected even if a large $N$ limit is assumed. Indeed, since the large $N$ limit is taken 
for $Ne_0^2$ fixed, it follows that $\gamma_A\sim{\cal O}(1)$, as it should be, 
since it gives the anomalous dimension RG function of the non-compact 
theory. It is of the same order as the first term 
between parentheses in Eq. (\ref{betay-new}), 
which corresponds to the dimensionality of the space-time. Thus, this problem has 
no obvious control parameter, and should be seen as a matter field fluctuation-corrected 
Debye-H\"uckel theory. In the case of compact Maxwell theory, the Debye-H\"uckel 
theory corresponds to a non-dilute gas of monopoles and its validity is determined the 
parameter $n\lambda_D^3$, where $n$ is the monopole density and 
$\lambda_D\equiv\sqrt{e^2/(4\pi^2 n)}$ is the Debye Length. For the compact Maxwell theory 
we have that $n\lambda_D^3\gg 1$ and the Debye-H\"uckel theory is a very good 
approximation. Including matter fields makes the monopole 
gas dilute and $n\lambda_D^3$ is no longer 
large. The Debye-H\"uckel parameter can be written as 
$n\lambda_D^3=\sqrt{2}e^2/(8\pi^2M)$, where $M=2\pi\sqrt{2z}/e$ is the photon mass. 
Thus, in the presence of matter a perturbation theory around the compact Maxwell 
theory can be performed where $e^2/M$ is a small parameter. We will see below that 
the fixed points at nonzero fugacity give indeed a small value of $n\lambda_D^3$.  

By considering the one-loop result $\gamma_A=Nf/8$, 
we find besides the fixed points $f_*=8/N$ and $y_*=0$ of the non-compact 
theory, the following non-trivial fixed points governing 
the phase structure of compact QED3: 
\begin{eqnarray}
\label{f*}
f_{\pm}&=&\frac{4}{N}\left(6\pm\sqrt{36-N\pi^3}\right),\\
\label{y*}
y_\pm&=&\frac{1}{\pi}\left(\frac{60-N\pi^3\pm10\sqrt{36-N\pi^3}}{2N}\right)^{1/2}.
\end{eqnarray}
The above fixed points exist only for $N<N_c=36/\pi^3\approx 1.161$. For 
$N>N_c$ only the non-compact fixed point exists. For $N=1$ the 
fixed points $(f_\pm,y_\pm)$ give for the Debye-H\"uckel parameter the values 
$(n\lambda_D^3)_+\approx 0.3$ and $(n\lambda_D^3)_-\approx 0.15$, respectively.     
In Fig. 1 we show a schematic flow diagram for the case $N=1$. The dashed line 
in the flow diagram passes through the {\it unstable} fixed point 
having coordinates $(f_-,y_-)$. This line separates two different critical 
regimes.   
\begin{figure}
\includegraphics[width=7cm]{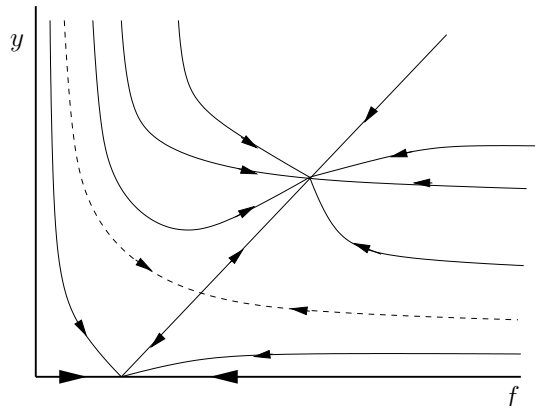}
\caption{Schematic flow diagram for the case $N=1$}
\end{figure}
Note that $\gamma_A$ is $N$-dependent at the fixed points $(f_\pm,y_\pm)$, 
with its two possible values given by 
$\gamma_A^{\pm}=Nf_\pm/8=(6\pm\sqrt{36-N\pi^3})/2$. 
%
{\it This result implies that 
there is no $N\geq 1$ for which $\gamma_A^\pm=1$ in 
compact QED3}. This rules out a KT-like transition in   
compact QED3 for physical values of $N$. 

The flow diagram 
in Fig. 1 indicates two distinct physical regimes governed by stable 
fixed points separated by the dashed line in the figure. 
Depending on the initial conditions on the physical parameters, 
the system will choose to flow either to the non-compact fixed point below 
the dashed line, or to the compact one above the dashed line.    
The interesting physical regime for us is governed 
by the fixed points at nonzero fugacity. It is 
clear that the fixed points $(f_\pm,y_\pm)$ are associated with confined 
phases, since there both the photon mass $M$ and string 
tension $\sigma=2e^2M/\pi^2$ are nonzero.   
The string tension approaches a universal value as   
$N$ approaches $N_c$ from the left, i.e., 
$\lim_{N\to N_c^-}\sigma/e_0^4=8(\pi/3)^{3/4}$. Since for $N>N_c$ the 
string tension vanishes, it follows that there is a universal jump at 
$N_c$. Thus, in the present context the string stiffness behaves similarly to 
the superfluid stiffness in two-dimensional superfluids \cite{NelKost}, though here 
there is no KT transition.  The vanishing of the string tension above $N_c$ 
is a clear signature for spinon deconfinement for $N=2$.  

Below $N_c$ the interspinon potential has the form 
$V(R)=\sigma R-\alpha/R+{\cal O}(1/R^2)$, where $\alpha$ is the universal coefficient of  
the L\"uscher term for the string fluctuation in compact QED3. The coefficient   
$\alpha$ is defined by $\alpha=f_c/2\pi$, where $f_c$ is any of the three 
charged fixed points in Fig. 1. For the stable confining 
regime governed by the fixed point $(f_+,y_+)$ we obtain that 
$\alpha=2(6+\sqrt{36-N\pi^3})/\pi N$.      

It is perfectly plausible to argue that in the confined 
phase of compact QED3 the chiral symmetry is broken, just as in the 
QCD case \cite{Casher}. Chiral symmetry breaking is 
believed to occur in QED3 for $N<N_{\rm ch}$, where typically 
$N_{\rm ch}\sim 3$. Indeed, an early estimate based 
on the analysis of the Schwinger-Dyson equation 
gives $N_{\rm ch}=32/\pi^2\approx 3.2$ \cite{Appelquist}, 
which was roughly confirmed by a Monte Carlo simulation giving 
$N_{\rm ch}=3.5\pm0.5$ \cite{Dagotto}. However, the true value of 
$N_{\rm ch}$ is still far from being consensual. 
For instance, recent Monte Carlo simulations 
do not find a decisive indication that chiral symmetry is broken 
for $N\geq 2$ \cite{Hands} and  
an elaborate analysis of 
the Schwinger-Dyson equations gives $N_{\rm ch}\approx4$ \cite{Fischer}.
Our results indicate that for 
$N_c<N<N_{\rm ch}$ the spinons are deconfined but chiral symmetry is broken. 
However, it is not excluded that $N_c=N_{\rm ch}$. 
Recently, a conjectured inequality was used to suggest that $N_{\rm ch}=3/2$ in 
the non-compact case  \cite{Appel}. It is remarkable that our critical value $N_c$ is 
so close to the latter estimate.  
If $N_{\rm ch}>2$, we would 
obtain that for the physical case $N=2$ antiferromagnetism is present \cite{Kim}, 
while the spinons are deconfined. In such a situation doping will eventually 
destroy the magnetic order and, since the spinons are deconfined, a genuine 
spin liquid will develop. Our results confirm 
the analysis of Ref. \cite{Hermele}, whose discussion 
was made in the large $N$ limit. 

We are indebted to Zlatko Tesanovic, who pointed out an important mistake 
in a previous version of the paper. This work received partial 
support of the european network COSLAB.

\end{document}